\documentclass[12pt]{amsart}
\usepackage{amssymb,latexsym,amsmath,enumitem,mathrsfs,geometry}
\usepackage{thmtools}
\usepackage{fullpage}
\usepackage{fancyhdr}
\usepackage{xcolor}
\usepackage{bbm}
\usepackage{stmaryrd}
\usepackage{mathrsfs}
\usepackage{hyperref}
\usepackage{eso-pic}
\usepackage{graphicx}
\usepackage{lineno}
\usepackage{diagbox}
\usepackage{array}
\usepackage{tikz}
\usetikzlibrary{arrows}
\usetikzlibrary{positioning, arrows.meta, shapes.geometric}
\usepackage{float}
\usepackage{comment}
\usepackage{mathtools}
\usepackage{amsthm}
\usepackage{placeins}
\usepackage{acro}
\usepackage{setspace}
\newtheorem{theorem}{Theorem}[section]
\newtheorem{definition}{Definition}[section]
\newtheorem{corollary}{Corollary}[section]
\newtheorem{proposition}{Proposition}[section]
\newtheorem{remark}{Remark}[section]
\newtheorem{example}{Example}[section]
\DeclareAcronym{RV}{
short = RV,
long  = \emph{Random Variable},
tag   = abbrev}
\DeclareAcronym{KL}{
short = KL,
long  = \emph{Kullback--Leibler},
tag   = abbrev}
\DeclareAcronym{GM}{
short = GM,
long  = \emph{Gaussian Mixture},
tag   = abbrev}
\DeclareAcronym{PDF}{
short = PDF,
long  = \emph{Probability Density Function},
tag   = abbrev}
\DeclareAcronym{EM}{
short = EM,
long  = \emph{Expectation-Maximization},
tag   = abbrev}
\usepackage{cleveref}
\usepackage[
    backend=biber,
    style=numeric,
    sorting=none,
    doi=false,
    url=false,
    eprint=false,
    isbn=false
]{biblatex}
\begin{document}
\numberwithin{equation}{section}
%%%%%%%%%%%%%%%%%%%%%%%%%%%%%%%%%%%%%%%%%%%%%%%%%%%%%%%%%%%%%%%%%%%%%%%%%%%%%%%%%%%%%
%%%%%%%%%%%%%%%%%%%%%%%%%%%%%%%%%%%%%%%%%%%%%%%%%%%%%%%%%%%%%%%%%%%%%%%%%%%%%%%%%%%%%
\title{Closed-Form Gaussian Mixture Representations of d-Dimensional Triangulated Domains}
%%%%%%%%%%%%%%%%%%%%%%%%%%%%%%%%%%%%%%%%%%%%%%%%%%%%%%%%%%%%%%%%%%%%%%%%%%%%%%%%%%%%%
%%%%%%%%%%%%%%%%%%%%%%%%%%%%%%%%%%%%%%%%%%%%%%%%%%%%%%%%%%%%%%%%%%%%%%%%%%%%%%%%%%%%%
\author{Ali Darijani$^{\circledast\square\#\P\perp}$}
\address{
Ali Darijani$^{\circledast\square\#\P\perp}$\newline
\indent Research Staff, KIT Department of Computer Science, Germany\newline
\indent Research Staff, Fraunhofer IOSB of Fraunhofer-Gesellschaft, Germany
}
\email{ali.darijani@rwth-aachen.de}
%%%%%%%%%%%%%%%%%%%%%%%%%%%%%%%%%%%%%%%%%%%%%%%%%%%%%%%%%%%%%%%%%%%%%%%%%%%%%%%%%%%%%
%%%%%%%%%%%%%%%%%%%%%%%%%%%%%%%%%%%%%%%%%%%%%%%%%%%%%%%%%%%%%%%%%%%%%%%%%%%%%%%%%%%%%
\author{Benedikt Stratmann$^{\#\P\perp}$}
\address{
Benedikt Stratmann$^{\#\P\perp}$\newline
\indent Research Staff, Fraunhofer IOSB of Fraunhofer-Gesellschaft, Germany}
\email{benedikt.stratmann@iosb.fraunhofer.de}
\author{Ratan Bahadur Thapa$^{\P\perp}$}
\address{
Ratan Bahadur Thapa$^{\P\perp}$\newline
\indent Research Staff, University of Stuttgart Department of Computer Science, Germany
}
\email{ratan.thapa@ki.uni-stuttgart.de}
%%%%%%%%%%%%%%%%%%%%%%%%%%%%%%%%%%%%%%%%%%%%%%%%%%%%%%%%%%%%%%%%%%%%%%%%%%%%%%%%%%%%%
%%%%%%%%%%%%%%%%%%%%%%%%%%%%%%%%%%%%%%%%%%%%%%%%%%%%%%%%%%%%%%%%%%%%%%%%%%%%%%%%%%%%%
\author{Uwe D.~Hanebeck$^{\circledast\P\perp}$}
\address{
Uwe D.~Hanebeck$^{\circledast\P\perp}$\newline
\indent Fellow, Institute of Electrical and Electronics Engineers (IEEE), United States\newline
\indent Professor, KIT Department of Computer Science, Germany
}
\email{uwe.hanebeck@ieee.org}
\email{uwe.hanebeck@kit.edu}
%%%%%%%%%%%%%%%%%%%%%%%%%%%%%%%%%%%%%%%%%%%%%%%%%%%%%%%%%%%%%%%%%%%%%%%%%%%%%%%%%%%%%
%%%%%%%%%%%%%%%%%%%%%%%%%%%%%%%%%%%%%%%%%%%%%%%%%%%%%%%%%%%%%%%%%%%%%%%%%%%%%%%%%%%%%
\author{Jürgen Beyerer$^{\circledast\perp}$}
\address{
Jürgen Beyerer$^{\circledast\perp}$\newline
\indent Professor, KIT Department of Computer Science, Germany\newline
\indent Director, Fraunhofer IOSB of Fraunhofer-Gesellschaft, Germany
}
\email{juergen.beyerer@kit.edu}
\email{juergen.beyerer@iosb.fraunhofer.de}
%%%%%%%%%%%%%%%%%%%%%%%%%%%%%%%%%%%%%%%%%%%%%%%%%%%%%%%%%%%%%%%%%%%%%%%%%%%%%%%%%%%%%
%%%%%%%%%%%%%%%%%%%%%%%%%%%%%%%%%%%%%%%%%%%%%%%%%%%%%%%%%%%%%%%%%%%%%%%%%%%%%%%%%%%%%
\author{Nehal Afifi$^{\circledast\perp}$}
\address{
Nehal Afifi$^{\circledast\perp}$\newline
\indent Research Staff, KIT Department of Mechanical Engineering, Germany
}
\email{nehal.afifi@kit.edu}
%%%%%%%%%%%%%%%%%%%%%%%%%%%%%%%%%%%%%%%%%%%%%%%%%%%%%%%%%%%%%%%%%%%%%%%%%%%%%%%%%%%%%
%%%%%%%%%%%%%%%%%%%%%%%%%%%%%%%%%%%%%%%%%%%%%%%%%%%%%%%%%%%%%%%%%%%%%%%%%%%%%%%%%%%%%
\address{
Symbol Legend\newline
\indent${\circledast}$: Idea, ${\square}$: Math, ${\#}$: Programming, ${\P}$: Writing, ${\perp}$: Critique
}
%%%%%%%%%%%%%%%%%%%%%%%%%%%%%%%%%%%%%%%%%%%%%%%%%%%%%%%%%%%%%%%%%%%%%%%%%%%%%%%%%%%%%
%%%%%%%%%%%%%%%%%%%%%%%%%%%%%%%%%%%%%%%%%%%%%%%%%%%%%%%%%%%%%%%%%%%%%%%%%%%%%%%%%%%%%
\keywords{Gaussian Mixture, Simplicial Complex}
%%%%%%%%%%%%%%%%%%%%%%%%%%%%%%%%%%%%%%%%%%%%%%%%%%%%%%%%%%%%%%%%%%%%%%%%%%%%%%%%%%%%%
%%%%%%%%%%%%%%%%%%%%%%%%%%%%%%%%%%%%%%%%%%%%%%%%%%%%%%%%%%%%%%%%%%%%%%%%%%%%%%%%%%%%%
\subjclass[2020]{94A15, 68T37, 62R40, 60B05}
%%%%%%%%%%%%%%%%%%%%%%%%%%%%%%%%%%%%%%%%%%%%%%%%%%%%%%%%%%%%%%%%%%%%%%%%%%%%%%%%%%%%%
%%%%%%%%%%%%%%%%%%%%%%%%%%%%%%%%%%%%%%%%%%%%%%%%%%%%%%%%%%%%%%%%%%%%%%%%%%%%%%%%%%%%%
\thanks{Funded by the Deutsche Forschungsgemeinschaft (DFG) - Project number 471687386}
%%%%%%%%%%%%%%%%%%%%%%%%%%%%%%%%%%%%%%%%%%%%%%%%%%%%%%%%%%%%%%%%%%%%%%%%%%%%%%%%%%%%%
%%%%%%%%%%%%%%%%%%%%%%%%%%%%%%%%%%%%%%%%%%%%%%%%%%%%%%%%%%%%%%%%%%%%%%%%%%%%%%%%%%%%%
\date{\today}
%%%%%%%%%%%%%%%%%%%%%%%%%%%%%%%%%%%%%%%%%%%%%%%%%%%%%%%%%%%%%%%%%%%%%%%%%%%%%%%%%%%%%
%%%%%%%%%%%%%%%%%%%%%%%%%%%%%%%%%%%%%%%%%%%%%%%%%%%%%%%%%%%%%%%%%%%%%%%%%%%%%%%%%%%%%
\begin{abstract}
We derive a Gaussian Mixture (GM) representation of a full-dimensional subset of Euclidean space with finite volume from a finite simplicial triangulation in closed form. On each simplex, we replace the uniform probability density function (PDF) with a Gaussian PDF that minimizes the Kullback--Leibler (KL) divergence, arriving at the information projection onto the family of Gaussian PDFs. By exploiting affine isomorphisms between simplices, the joint convexity and bijection invariance properties of the KL divergence, identities for Dirichlet PDFs, and a partition-of-unity argument, we establish a dimension-dependent but geometry-independent sharp upper bound on the global KL divergence between the resulting GM PDF and the uniform PDF on the original domain. Numerical examples illustrate the construction and demonstrate its potential in practical applications.
\end{abstract}
%%%%%%%%%%%%%%%%%%%%%%%%%%%%%%%%%%%%%%%%%%%%%%%%%%%%%%%%%%%%%%%%%%%%%%%%%%%%%%%%%%%%%
%%%%%%%%%%%%%%%%%%%%%%%%%%%%%%%%%%%%%%%%%%%%%%%%%%%%%%%%%%%%%%%%%%%%%%%%%%%%%%%%%%%%%
\maketitle
\AddToShipoutPictureFG{%
  \AtPageCenter{%
    \put(0,0){%
      \makebox(0,0)[l]{%
        \hspace*{0.40\paperwidth}%
        \rotatebox{-90}{%
          \parbox{0.74\paperheight}{%
            \normalsize
            \normalfont
            \upshape
            \copyright\ 2026 IEEE. Personal use of this material
            is permitted. Permission from IEEE must be obtained
            for all other uses, in any current or future media,
            including reprinting/republishing this material for
            advertising or promotional purposes, creating new
            collective works, for resale or redistribution to
            servers or lists, or reuse of any copyrighted
            component of this work in other works.
          }%
        }%
      }%
    }%
  }%
}
%%%%%%%%%%%%%%%%%%%%%%%%%%%%%%%%%%%%%%%%%%%%%%%%%%%%%%%%%%%%%%%%%%%%%%%%%%%%%%%%%%%%%
%%%%%%%%%%%%%%%%%%%%%%%%%%%%%%%%%%%%%%%%%%%%%%%%%%%%%%%%%%%%%%%%%%%%%%%%%%%%%%%%%%%%%
\section{Introduction}
\label{sec:introduction}

Many engineering and physical systems are geometric~\cite{marsden_geometric_mechanics,marsden_symmetry_mechanics,arnold_mathematical_methods_of_classical_mechanics,frankel_geometry_of_physics} before they are
probabilistic. Objects are defined by their shape, environments by spatial
maps, and mechanical systems by admissible configurations. At the same time,
uncertainty is unavoidable due to noisy measurements, incomplete knowledge,
and variability. A natural question is therefore how a known geometric
description can be converted directly into a probabilistic representation
suitable for estimation, fusion, and decision-making.

\acp{GM} provide a flexible representation of
probability densities and positive measures and can be interpreted as
basis functions for such representations~\cite{mazya_schmidt_approximate_approximations,stuart_gaussian_mixture_approximation}. However,
most existing \ac{GM} construction methods solve a different problem: they infer
a mixture from samples or a discretized representation of an unknown
distribution. \ac{EM}~\cite{maximum_likelihood_from_incomplete_data_via_the_em_algorithm_dempster_laird_rubin,em_convergence_wu,em_algorithm_mclachlan_krishnan},
Bayesian approaches, and grid-based density representations are powerful in
this setting, but require additional choices concerning samples or grids,
initialization, optimization, stopping criteria, and the number of mixture
components. Their iterative or discretization-dependent nature also makes it
difficult to obtain a direct geometry-to-probability mapping together with an
analytical approximation guarantee in an appropriate probabilistic distance measure. In particular, existence results for
\ac{GM} approximations do not provide a constructive procedure for obtaining the
parameters from a given geometric object~\cite{mazya_schmidt_approximate_approximations,stuart_gaussian_mixture_approximation}.

\begin{figure}[t]
    \centering
    \includegraphics[width=\linewidth]{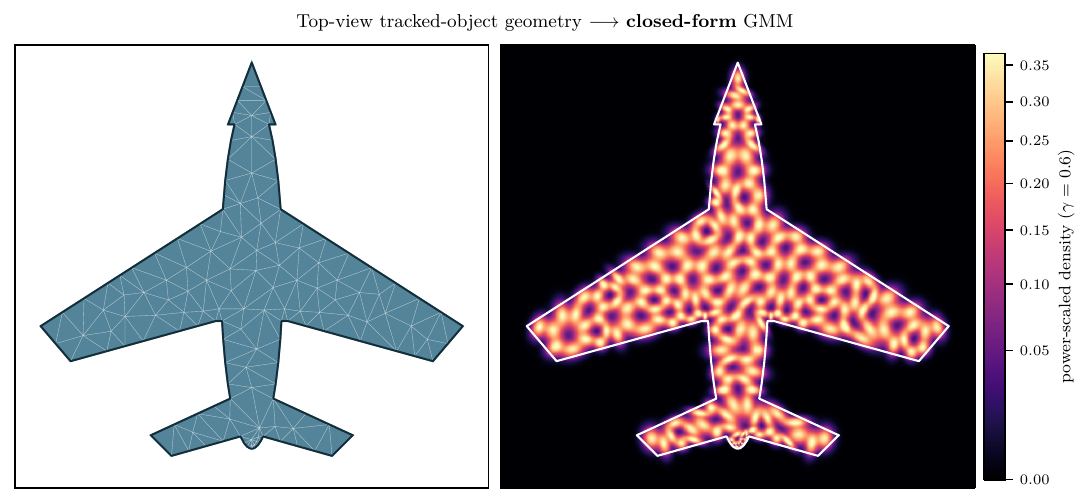}
    \caption{A fixed-wing planar silhouette is triangulated and directly
    converted into a \ac{GM}, with one closed-form component per
    triangle.}
    \label{fig:plane-teaser}
\end{figure}

This work considers the complementary, geometry-driven problem: the target \ac{PDF} is known over a geometric domain, and the objective is to
construct its \ac{GM} representation directly. Given a uniform \ac{PDF} over a finite triangulated domain, where each ``triangle'' is the full-dimensional analogue of a tetrahedron in higher dimensions, we seek a reasonably accurate \ac{GM} representation. We replace this uniform distribution locally by its Gaussian I-projection and
combine the resulting components according to the ``triangle'' volumes. The
Gaussian parameters are obtained directly from the ``triangle'' vertices, so the
complete \ac{GM} is constructed in closed form without sampling or iterative
optimization. The topological equivalence of ``triangles'' further allows the local
approximation error to be characterized independently of the position,
scale, and shape of the ``triangles'' within the triangulation.~\Cref{fig:plane-teaser} illustrates this direct construction for a planar
fixed-wing silhouette.

The resulting representation is particularly useful when geometry is part of
the computational model. For example, geometric refinement naturally produces
a finer probabilistic representation: in the experiments, barycentric and
polyhedral refinement empirically reduce the global forward \ac{KL}
divergence. At equal component counts, the geometry-derived construction also
avoids the iterative fitting required by \ac{EM}. In a nut--bolt clearance
example, translations of the geometric object directly translate the
corresponding Gaussian means, enabling analytic overlap rescoring without
refitting the \ac{GM}. In a map-aware fusion example, the geometry-derived
prior can be combined sequentially with Gaussian sensor likelihoods in
closed form. These examples illustrate that the \ac{GM} is not merely a
statistical approximation, but retains an explicit correspondence to the
underlying geometric representation.

The approach is also motivated by the close relationship between geometry and
physical modeling. Geometry provides the structure of physical systems,
including configurations, constraints, and admissible states, while
probability provides a natural representation of uncertainty. Geometric
formulations of mechanics and physics illustrate the fundamental role of
geometry in describing physical systems~\cite{marsden_geometric_mechanics,marsden_symmetry_mechanics,arnold_mathematical_methods_of_classical_mechanics,frankel_geometry_of_physics}.
Combining this geometric structure with probabilistic inference and
uncertainty-aware control~\cite{fleming_rishel_deterministic_and_stochastic_optimal_control} suggests a broader
``probabilistic physics'' perspective in which geometry, probability, and
decision making share a common computational representation.

The main contribution of this work is therefore an explicit
geometry-to-\ac{GM} construction for uniform \ac{PDF} on finite
full-dimensional triangulated domains. The construction is closed form, uses
one Gaussian component per ``triangle'', and requires no numerical optimization.
Furthermore, the resulting mixture admits a dimension-dependent sharp upper bound
on the global forward \ac{KL} divergence. The number of components is
controlled by the chosen triangulation, providing a natural accuracy--complexity
trade-off. Refinement of the triangulation and its corresponding \ac{KL} error suggests further investigation of
adaptive triangulation and component reduction. The remainder of this paper
develops the construction, establishes the local and global approximation
results, and evaluates the resulting representation through geometric
refinement, clearance analysis, and probabilistic fusion.
%%%%%%%%%%%%%%%%%%%%%%%%%%%%%%%%%%%%%%%%%%%%%%%%%%%%%%%%%%%%%%%%%%%%%%%%%%%%%%%%%%%%%
%%%%%%%%%%%%%%%%%%%%%%%%%%%%%%%%%%%%%%%%%%%%%%%%%%%%%%%%%%%%%%%%%%%%%%%%%%%%%%%%%%%%%
\section{Construction}
\begin{definition}[Convex Hull~\cite{elements_of_algebraic_topology_munkres_krantz_parks}]
Let $A \subset \mathbb{R}^d$. The convex hull of $A$ is defined by:
\begin{equation}
\operatorname{conv}(A)
\coloneq
\left\{
\sum_{i=1}^n \lambda_i v_i \colon
n \in \mathbb{N}_{>0},\;
v_i \in A,\;
\lambda_i \ge 0,\;
\sum_{i=1}^n \lambda_i = 1
\right\}.
\end{equation}
\end{definition}

\begin{definition}[Standard Simplex~\cite{elements_of_algebraic_topology_munkres_krantz_parks}]
The standard $d$-simplex $\Delta^d \subset \mathbb{R}^d$ is defined by:
\begin{equation}
\Delta^d \coloneq \operatorname{conv}\{0,e_1,\dots,e_d\},
\end{equation}
where $e_i$ denotes the $i$-th canonical basis in $\mathbb{R}^d$.
\end{definition}

\begin{definition}[Geometrically Independent Points~\cite{elements_of_algebraic_topology_munkres_krantz_parks}]
$\{v_i\}_{i=0,\dots,d} \subset \mathbb{R}^d$ are geometrically independent iff $\{v_i - v_0\}_{i=1,\dots,d}$ are linearly independent.
\end{definition}

\begin{definition}[General Simplex~\cite{elements_of_algebraic_topology_munkres_krantz_parks}]
Let $v_0,\dots,v_d \in \mathbb{R}^d$ be geometrically independent. The general $d$-simplex $\sigma^d$ generated by these points is defined by
\begin{equation}
\sigma^d(v_0,\dots,v_d) \coloneq \operatorname{conv}\{v_0,\dots,v_d\}.
\end{equation}
\end{definition}

\begin{definition}[Faces of a Simplex~\cite{elements_of_algebraic_topology_munkres_krantz_parks}]
Let
\[
\sigma^d=\operatorname{conv}\{v_0,\dots,v_d\}
\]
be a \(d\)-simplex. A \emph{face} of \(\sigma^d\) is a simplex of the form
\[
\tau\coloneq\operatorname{conv}\{v_i:i\in I\},
\qquad \varnothing\neq I\subseteq\{0,\dots,d\}.
\]
If \(|I|=r+1\), then \(\dim\tau=r\). In particular, \(0\)-faces, \(1\)-faces, and \((d-1)\)-faces are called \emph{vertices}, \emph{edges}, and \emph{facets}, respectively.
\end{definition}

\begin{definition}[Geometric Simplicial Complex~\cite{elements_of_algebraic_topology_munkres_krantz_parks}]
\label{def:geometric-simplicial-complex}
A \emph{geometric simplicial complex} $\mathcal{S}$ in
$\mathbb{R}^d$ is a collection of simplices in $\mathbb{R}^d$ such that:
\begin{enumerate}
    \item every face of a simplex in $\mathcal{S}$ is also in $\mathcal{S}$;
    \item for any $\sigma,\tau\in\mathcal{S}$,
    \[
    \sigma\cap\tau
    \]
    is either empty or a common face of both $\sigma$ and $\tau$.
\end{enumerate}
\end{definition}

\begin{definition}[Geometric Realization~\cite{kozlov_combinatorial_algebraic_topology}]
The \emph{geometric realization} of a geometric simplicial complex \(\mathcal{S}\) is
\[
|\mathcal{S}|
\coloneqq
\bigcup_{\sigma\in\mathcal{S}}\sigma
\subseteq \mathbb{R}^d.
\]
\end{definition}

\begin{definition}[Dimension and Purity~\cite{elements_of_algebraic_topology_munkres_krantz_parks}]
Let $\mathcal{S}$ be a finite geometric simplicial complex. Its
\emph{dimension} is
\[
\dim\mathcal{S}
\coloneqq
\max_{\sigma\in\mathcal{S}}\dim\sigma.
\]
A simplex $\sigma\in\mathcal{S}$ is \emph{maximal} (or a
\emph{facet}) if it is not a proper face of any simplex
$\tau\in\mathcal{S}$. The complex $\mathcal{S}$ is
\emph{pure of dimension $d$} if every maximal simplex has
dimension $d$.
\end{definition}

\begin{definition}[Geometric Triangulation~\cite{elements_of_algebraic_topology_munkres_krantz_parks}] \label{def:triangulation} Let \[ \Omega\subset\mathbb{R}^d, \qquad \dim\Omega=d, \qquad \operatorname{vol}_d(\Omega)<+\infty. \] A \emph{finite geometric triangulation} of $\Omega$ is a pure $d$-dimensional geometric simplicial complex $\mathcal{S}$ such that \[ |\mathcal{S}|=\Omega, \qquad \#\mathcal{S}<\infty. \] We denote its set of $d$-simplices by \[ \mathcal{S}^{[d]} \coloneqq \{\sigma\in\mathcal{S}:\dim\sigma=d\}. \] \end{definition}

\begin{theorem}[Triangulation for 2-Manifolds~\cite{lee_introduction_to_topological_manifolds}]
    Every $2$-manifold is homeomorphic to the polyhedron of a $2$-dimensional simplicial complex, in which every $1$-simplex is a face of exactly two $2$-simplices.
\end{theorem}
\begin{theorem}[Triangulation for 3-Manifolds~\cite{lee_introduction_to_topological_manifolds}]
Every 3-manifold is triangulable.
\end{theorem}

\begin{theorem}[Triangulation for Differentiable Manifolds~\cite{on_c1_complexes_whitehead}]
    Differentiable manifolds are always triangulable.
\end{theorem}

\begin{proposition}[Intersections of Maximal Simplices~\cite{tao_an_introduction_to_measure_theory,elements_of_algebraic_topology_munkres_krantz_parks}]
\label{prop:maximal-simplex-intersection}
Let \(\mathcal{S}\) be a finite, pure, \(d\)-dimensional geometric simplicial complex. Then, for distinct \(\sigma,\tau\in\mathcal{S}^{[d]}\),
\[
\sigma\cap\tau=\varnothing
\quad\text{or}\quad
\sigma\cap\tau\in\mathcal{S},
\qquad
\dim(\sigma\cap\tau)\le d-1.
\]
Hence
\[
\operatorname{vol}_d(\sigma\cap\tau)=0,
\]
and \(\operatorname{int}(\sigma)\cap\operatorname{int}(\tau)=\varnothing\).
\end{proposition}

\begin{proof}
By the definition of a geometric simplicial complex, \(\sigma\cap\tau\) is empty or a common face. Since \(\sigma\ne\tau\) and \(\dim\sigma=\dim\tau=d\), any non-empty common face has dimension at most \(d-1\), hence has \(d\)-dimensional Lebesgue measure zero.
\end{proof}

\begin{corollary}[Volume of a Finite triangulation~\cite{tao_an_introduction_to_measure_theory}]
\label{cor:volume-triangulation}
Let \(\mathcal{S}\) be a finite, pure, \(d\)-dimensional geometric simplicial complex. Then
\[
\operatorname{vol}_d(|\mathcal{S}|)
=
\sum_{\sigma\in\mathcal{S}^{[d]}}
\operatorname{vol}_d(\sigma).
\]
\end{corollary}

\begin{proof}
Since
\[
|\mathcal{S}|=\bigcup_{\sigma\in\mathcal{S}^{[d]}}\sigma
\]
and distinct \(d\)-simplices intersect only in sets of \(d\)-dimensional measure zero, finite additivity yields the result.
\end{proof}

\begin{proposition}[$F:\Delta^d \stackrel{\mathrm{aff}}{\cong} \sigma^d(v_0,\dots,v_d)$]

Let $\Delta^d$ be the standard $d$-simplex, and\\ $\sigma^d(v_0,\ldots,v_d)
=\operatorname{conv}\{v_0,\ldots,v_d\}
$ be the general $d$-simplex, Then there exists a unique affine isomorphism
\begin{equation}
F:\Delta^d\overset{\mathrm{aff}}{\cong}\sigma^d(v_0,\ldots,v_d), \quad  x \mapsto Ax+v_0,
\end{equation}

where
\begin{equation}
A=
\begin{bmatrix}
v_1-v_0 & v_2-v_0 & \cdots & v_d-v_0
\end{bmatrix}.   
\end{equation}
\end{proposition}

\begin{proposition}[$\operatorname{vol}_d\sigma(v_0,\dots,v_d)$~\cite{stein_simplex_volume,dirichlet_and_related_distributions}]
\label{prop:simplex-volume}
\begin{equation}
\operatorname{vol}_d\sigma(v_0,\dots,v_d)
=
\frac{1}{d!}
\left(
\det G
\right)^{1/2},
\end{equation}
where $G$ is the Gram matrix: 
\begin{equation}
G_{ij} \coloneq (v_i - v_0)^\top (v_j - v_0), \quad i,j=1 \dots,d.
\end{equation}
\end{proposition}

\begin{definition}[Dirichlet Distribution on $\Delta^d$~\cite{dirichlet_and_related_distributions}]
Let $\Delta^d = \operatorname{conv}\{0,e_1,\ldots,e_d\}$. For
$x=(x_1,\ldots,x_d)\in\Delta^d$, define
\[
x_0 := 1-\sum_{i=1}^d x_i.
\]
Thus $(x_0,\ldots,x_d)$ are the barycentric coordinates of $x$.
For $\alpha_{\sharp}=(\alpha_0,\ldots,\alpha_d)\in\mathbb{R}_{++}^{d+1}$, the
Dirichlet density on $\Delta^d$ is
\[
\mathcal{D}_{\Delta^d}(x|\alpha_\sharp)
=
\frac{1}{B(\alpha_\sharp)}
\prod_{i=0}^d x_i^{\alpha_i-1},
\]
where
\[
B(\alpha_\sharp)
=
\frac{\prod_{i=0}^d \Gamma(\alpha_i)}
{\Gamma(|\alpha_\sharp|)},
\]
\end{definition}

\begin{corollary}[Uniform \acs{PDF} on $\Delta^d$ in $\mathbb{R}^d$~\cite{stein_simplex_volume,dirichlet_and_related_distributions}]
\label{cor:uniform-pdf-simplex}
Let $1_{d+1} \coloneqq (1,\ldots,1)\in \mathbb{R}^{d+1}$.  
The uniform probability density function on $\Delta^d$ is the Dirichlet density with parameter
$1_{d+1}$, i.e.
\begin{equation}
X \sim \mathcal{D}_{\Delta^d}(x|1_{d+1}).
\label{eq:uniform-dirichlet-notation}
\end{equation}
Its density is given by
\begin{equation}
\mathcal{U}_{\Delta^d}(x)
\coloneqq
\mathcal{D}_{\Delta^d}(x|1_{d+1})
=
\begin{cases}
d!
&
x\in \Delta^d
\\[6pt]
0
&
x\notin \Delta^d.
\end{cases}
\label{eq:uniform-pdf-simplex}
\end{equation}
\end{corollary}

\begin{proposition}[$\mathbb{E}(X),\mathbb{V}(X)$ for $X \sim \mathcal{D}_{\Delta^d}
\label{prop:dirichlet_mean_var}
(x|\alpha)$~\cite{dirichlet_and_related_distributions}]
Let $X=(X_1,\ldots,X_d)$ have a Dirichlet distribution on $\Delta^d$
with parameter $\alpha_{\sharp}=(\alpha_0,\ldots,\alpha_d)\in\mathbb{R}_{++}^{d+1}$ and $\alpha_{\natural}=(\alpha_1,\ldots,\alpha_d)\in\mathbb{R}_{++}^{d}$.
Then
\begin{equation}
\begin{split}    
\mathbb{E}(X)&=\frac{\alpha_{\natural}}{|\alpha_{\sharp}|},\\
\mathbb{V}(X)&=\frac{|\alpha_{\sharp}|\operatorname{diag}(\alpha_{\natural})-\alpha_{\natural}\alpha_{\natural}^{\top}}{|\alpha_{\sharp}|^2(|\alpha_{\sharp}|+1)}.
\end{split}
\label{eq:dirichlet-expectation}
\end{equation}
\end{proposition}

\begin{definition}[\ac{KL} Divergence]
\label{def:kl_divergence}
Let $P$ and $Q$ be probability measures on $\Omega=\mathbb{R}^d$. The
Kullback--Leibler (KL) divergence of $P$ from $Q$ is defined as:
\begin{equation}
D_{\mathrm{KL}}(P:Q)
\coloneq
\begin{cases}
\displaystyle
\int_{\Omega}
\log\left(\frac{\mathrm{d}P}{\mathrm{d}Q}\right)
\mathrm{d}P,
& P \ll Q,\\[8pt]
+\infty,
& \text{otherwise},
\end{cases}
\end{equation}
where $P \ll Q$ denotes that $P$ is absolutely continuous with respect
to $Q$.
\end{definition}

\begin{remark}[\ac{KL} Divergence for \acp{PDF}]
Suppose that $P$ and $Q$ admit densities $p$ and $q$ with respect to
the Lebesgue measure, then
\begin{equation}
D_{\mathrm{KL}}(P:Q)
=
\begin{cases}
\displaystyle
\int_{\Omega}
p(x)\log\left(\frac{p(x)}{q(x)}\right)\mathrm{d}x,
& P \ll Q,\\[8pt]
+\infty,
& \text{otherwise}.
\end{cases}
\end{equation}
\end{remark}

\begin{definition}[I-Projection]
Let $\mathcal{Q}$ be a non-empty set of \acp{PDF} and let $p$ be a \ac{PDF}. The I-projection of $p$ onto $\mathcal{Q}$ is defined as
\begin{equation}
\Pi_{\mathcal{Q}}^{\mathrm{I}}(p)
\coloneqq
\underset{q\in\mathcal{Q}}{\operatorname{arg\,min}}\,
D_{\mathrm{KL}}(p:q).
\end{equation}
\end{definition}

\begin{remark}
For example, the I-projection of a \ac{PDF} $p(x)$ onto the Gaussian family
$\mathcal{Q}_{\mathcal{N}} \coloneqq
\{\mathcal{N}(x|m,\Sigma):m\in\mathbb{R}^d,
\Sigma\in\mathbb{S}_{++}^d\}$ is the Gaussian \ac{PDF} that minimizes
$D_{\mathrm{KL}}\bigl(p(x):\mathcal{N}(x|m,\Sigma)\bigr)$ over
$m\in\mathbb{R}^d$ and $\Sigma\in\mathbb{S}_{++}^d$.
\end{remark}

\begin{definition}[M-Projection]
Let $\mathcal{Q}$ be a non-empty set of \acp{PDF} and let $p$ be a \ac{PDF}. The M-projection of $p$ onto $\mathcal{Q}$ is defined as
\begin{equation}
\Pi_{\mathcal{Q}}^{\mathrm{M}}(p)
\coloneqq
\underset{q\in\mathcal{Q}}{\operatorname{arg\,min}}\,
D_{\mathrm{KL}}(q:p).
\end{equation}
\end{definition}

\begin{remark}
For example, the M-projection of a \ac{PDF} $p(x)$ onto the Gaussian family
$\mathcal{Q}_{\mathcal{N}} \coloneqq
\{\mathcal{N}(x|m,\Sigma):m\in\mathbb{R}^d,
\Sigma\in\mathbb{S}_{++}^d\}$ is the Gaussian \ac{PDF} that minimizes
$D_{\mathrm{KL}}\bigl(\mathcal{N}(x|m,\Sigma):p(x)\bigr)$ over
$m\in\mathbb{R}^d$ and $\Sigma\in\mathbb{S}_{++}^d$.
\end{remark}

\begin{proposition}[$\min_{m\in\mathbb{R}^d,\Sigma\in\mathbb{S}_{++}^d}
D_{\mathrm{KL}}\left(
\mathcal{U}_{\Delta^d}(x):\mathcal{N}(x|m,\Sigma)
\right)$]
\label{prop:min-kl-uniform-standard-simplex-to-gaussian}
\begin{equation}
\begin{split}
\min_{m\in\mathbb{R}^d,\ \Sigma\in\mathbb{S}_{++}^d}
D_{\mathrm{KL}}\left(
\mathcal{U}_{\Delta^d}(x):\mathcal{N}(x|m,\Sigma)
\right)=
\log\left(
\frac{(2\pi e)^{d/2} (d!)}
{(d+1)^{(d+1)/2}(d+2)^{d/2}}\right).
\end{split}
\end{equation}
\begin{equation}
\label{eq:i_gaussian_projection}
\begin{split}
(m^\star,\Sigma^\star)&=\operatorname*{argmin}_{m\in\mathbb{R}^d,\Sigma\in\mathbb{S}_{++}^d}D_{\mathrm{KL}}\left(\mathcal{U}_{\Delta^d}(x):\mathcal{N}(x|m,\Sigma)\right)=\left(\mathbb{E}_{\mathcal{U}_{\Delta^d}}(X),\mathbb{V}_{\mathcal{U}_{\Delta^d}(x)}(X)\right).
\end{split}
\end{equation}

\end{proposition}

\begin{proof}
The derivation of \Cref{eq:i_gaussian_projection} follows from \cite{stuart_inverse_assimilation}. Using~\Cref{prop:dirichlet_mean_var} we have:
\begin{equation}
\begin{split}
(m^\star,\Sigma^\star)=\left(\frac1{d+1}1_d,\frac{(d+1)I_d-1_d 1_d^\top}{(d+1)^2(d+2)}\right), \quad\det\Sigma^\star=\frac1{(d+1)^{d+1}(d+2)^d}.
\end{split}
\end{equation}
Expanding~\Cref{def:kl_divergence} gives:
\begin{equation}
\begin{split}
\min_{m\in\mathbb R^d,\ \Sigma\in\mathbb S_{++}^d}D_{\mathrm{KL}}\left(\mathcal U_{\Delta^d}(x):\mathcal N(x| m,\Sigma)\right)=\log\left(\frac{(2\pi e)^{d/2}(d!)}{(d+1)^{(d+1)/2}(d+2)^{d/2}}\right).
\end{split}
\end{equation}
\end{proof}

\begin{proposition}[$D_{\mathrm{KL}}(p(x):q(x)$ is $\stackrel{\mathrm{aff}}{\cong}$ Invariant~\cite{information_geometry_ay_jost_le_schwachhofer}]
\label{prop:bijection_kl}
Let $p(x)$ and $q(x)$ be \acp{PDF} on $\mathbb{R}^d$, and let
\begin{equation}
f \colon \mathbb{R}^d \to \mathbb{R}^d, \quad x \mapsto y,
\end{equation}
be a diffeomorphism. Let $p'(y)$ and $q'(y)$ denote the \acp{PDF} obtained by transforming $p(x)$ and $q(x)$ under $f$, respectively.
Then
\begin{equation}
D_{\mathrm{KL}}\bigl(p'(y):q'(y)\bigr)
=
D_{\mathrm{KL}}\bigl(p(x):q(x)\bigr).
\end{equation}
\end{proposition}

\begin{proposition}[$\min_{m\in\mathbb{R}^d,\Sigma\in\mathbb{S}_{++}^d}
D_{\mathrm{KL}}\left(
\mathcal{U}_{\sigma^d}(y):\mathcal{N}(y|m,\Sigma)
\right)$]
\label{prop:min-kl-affine-uniform-d-simplex-to-gaussian}
For $\Delta^d$ and $\sigma^d(v_0,\ldots,v_d)$ let:
\begin{equation}
F:\Delta^d\overset{\mathrm{aff}}{\cong}\sigma^d(v_0,\ldots,v_d), \quad  x \mapsto Ax+v_0=y,
\end{equation}
be the affine diffeomorphism, then:
\begin{equation}
\begin{split}
\min_{m\in\mathbb{R}^d,\ \Sigma\in\mathbb{S}_{++}^d}
D_{\mathrm{KL}}\left(
\mathcal{U}_{\sigma^d}(y):
\mathcal{N}(y|m,\Sigma)
\right)=\log\left(
\frac{(2\pi e)^{d/2} (d!)}
{(d+1)^{(d+1)/2}(d+2)^{d/2}}
\right),
\end{split}
\end{equation}
and:
\begin{equation}
\begin{split}
(m^\star,\Sigma^\star)
=
\operatorname*{argmin}_{m\in\mathbb{R}^d,\Sigma\in\mathbb{S}_{++}^d}
D_{\mathrm{KL}}\left(
\mathcal{U}_{\sigma^d}(y):
\mathcal{N}(y|m,\Sigma)
\right)=
\left(
A\mathbb{E}_{\mathcal{U}_{\Delta^d}}(X)+v_0,
A\mathbb{V}_{\mathcal{U}_{\Delta^d}}(X)A^\top
\right).
\end{split}
\end{equation}
\end{proposition}

\begin{example}[$2$-simplex and its Gaussian I-projection]
\Cref{fig:kl-validation} numerically calculates and analyzes the analytical simplex \ac{KL} value in \Cref{prop:min-kl-affine-uniform-d-simplex-to-gaussian}.
\end{example}

\begin{remark}[I-projection vs M-projection]
In our setting, we consider the I-projection rather than the M-projection because the corresponding optimization problems have fundamentally different behavior. The I-projection attains a finite minimum,
\begin{equation}
\min_{q\in\mathcal{Q}}D_{\mathrm{KL}}(p:q)<\infty,
\end{equation}
whereas the reverse divergence appearing in the M-projection is infinite for every $q\in\mathcal{Q}$,
\begin{equation}
D_{\mathrm{KL}}(q:p)=\infty.
\end{equation}
Consequently, the M-projection does not provide a meaningful finite minimizer in this setting, while the I-projection does.
\end{remark}

\begin{proposition}[Uniform Measure as a Mixture]
\label{prop:uniform-mixture-triangulation}
Let \(\mathcal{S}\) be a finite geometric triangulation of \(\Omega\subseteq\mathbb{R}^d\land\dim\Omega=d\land0<\operatorname{vol}_d(\Omega)<+\infty\). Then
\[
\mathcal{U}_\Omega
=
\sum_{\sigma\in\mathcal{S}^{[d]}}
\frac{\operatorname{vol}_d(\sigma)}
     {\operatorname{vol}_d(\Omega)}
\,\mathcal{U}_\sigma
\qquad\text{a.e.},
\]
where
\[
\mathcal{U}_A(x)
\coloneqq
\frac{1_A(x)}{\operatorname{vol}_d(A)}.
\]
In particular, with
\[
w_\sigma
\coloneqq
\frac{\operatorname{vol}_d(\sigma)}
     {\operatorname{vol}_d(\Omega)},
\]
one has \(w_\sigma>0\) and
\[
\sum_{\sigma\in\mathcal{S}^{[d]}}w_\sigma=1.
\]
\end{proposition}

\begin{proof}
By \Cref{prop:maximal-simplex-intersection},
\[
1_\Omega
=
\sum_{\sigma\in\mathcal{S}^{[d]}}1_\sigma
\qquad\text{a.e.},
\]
and by \Cref{cor:volume-triangulation},
\[
\operatorname{vol}_d(\Omega)
=
\sum_{\sigma\in\mathcal{S}^{[d]}}\operatorname{vol}_d(\sigma).
\]
Therefore
\[
\sum_{\sigma\in\mathcal{S}^{[d]}}
w_\sigma\mathcal{U}_\sigma
=
\frac{1}{\operatorname{vol}_d(\Omega)}
\sum_{\sigma\in\mathcal{S}^{[d]}}1_\sigma
=
\mathcal{U}_\Omega
\qquad\text{a.e.},
\]
and \(\sum_\sigma w_\sigma=1\).
\end{proof}

\begin{proposition}[Joint Convexity of $D_{\mathrm{KL}}(p:q)$~\cite{information_geometry_ay_jost_le_schwachhofer}]

For any \acp{PDF} $p_1,p_2,q_1,q_2$ and any \(\lambda\in[0,1]\), we have:
\begin{equation}
\begin{split}
D_{\mathrm{KL}}\!\left(\lambda p_1+(1-\lambda)p_2 : \lambda q_1+(1-\lambda)q_2\right)\leq\lambda D_{\mathrm{KL}}(p_1:q_1)+(1-\lambda)D_{\mathrm{KL}}(p_2:q_2).
\end{split}
\end{equation}
\end{proposition}

\begin{proposition}
For \acp{PDF} $\{p_i\}_{i=1,\dots,K}$ and $\{q_i\}_{i=1,\dots,K}$, and $\pi \in \mathbb{R}^K_+ \land |\pi|_1=1$:
\begin{equation}
D_{\mathrm{KL}}(\sum_{i=1}^{K}\pi_i p_i:\sum_{i=1}^{K}\pi_i q_i)\le\sum_{i=1}^{K}\pi_iD_{\mathrm{KL}}(p_i:q_i).
\end{equation}
\end{proposition}

\begin{theorem}[$D_{\mathrm{KL}}\bigl(\mathcal{U}_\Omega(x):\mathcal{G}_{\mathcal{S}}(x|g)\bigr)\leq C(d)$]
\label{thm:main}
Let \(\Omega\subseteq\mathbb{R}^d\land\dim\Omega=d\land0<\operatorname{vol}_d(\Omega)<+\infty\) admit a finite geometric triangulation \(\mathcal{S}\). For each \(\sigma=\operatorname{conv}\{v_0,\dots,v_d\}\in\mathcal{S}^{[d]}\), define
\begin{equation}
    \begin{split}
 w_\sigma\coloneqq\frac{\operatorname{vol}_d(\sigma)}{\operatorname{vol}_d(\Omega)},\quad(m_\sigma^\star,\Sigma_\sigma^\star)
=
\underset{m\in\mathbb{R}^d,\;\Sigma\in\mathbb{S}_{++}^d}{\operatorname{argmin}}
\,D_{\mathrm{KL}}\bigl(\mathcal{U}_\sigma(x):\mathcal{N}(x| m,\Sigma)\bigr).      
    \end{split}
\end{equation}

Then
\[
\mathcal{G}_{\mathcal{S}}(x| g)
\coloneqq
\sum_{\sigma\in\mathcal{S}^{[d]}}
w_\sigma\mathcal{N}(x|m_\sigma^\star,\Sigma_\sigma^\star)
\]
is a \ac{GM} with \(\#\mathcal{S}^{[d]}\) components satisfying
\[
D_{\mathrm{KL}}\bigl(\mathcal{U}_\Omega(x):\mathcal{G}_{\mathcal{S}}(x|g)\bigr)
\le
\log\left(
\frac{(2\pi e)^{d/2}d!}
{(d+1)^{(d+1)/2}(d+2)^{d/2}}
\right).
\]
\end{theorem}

\begin{proof}
By \Cref{prop:uniform-mixture-triangulation},
\(\mathcal{U}_\Omega=\sum_{\sigma\in\mathcal{S}^{[d]}}w_\sigma\mathcal{U}_\sigma\) a.e.; hence, by the joint convexity of \(D_{\mathrm{KL}}\),
\[
D_{\mathrm{KL}}\bigl(\mathcal{U}_\Omega(x):\mathcal{G}_{\mathcal{S}}(x| g)\bigr)
\le
\sum_{\sigma\in\mathcal{S}^{[d]}}
w_\sigma
D_{\mathrm{KL}}
\bigl(
\mathcal{U}_\sigma:
\mathcal{N}(x|m_\sigma^\star,\Sigma_\sigma^\star)
\bigr).
\]
By \Cref{prop:min-kl-affine-uniform-d-simplex-to-gaussian},
\begin{equation}
    \begin{split}
 D_{\mathrm{KL}}
\bigl(
\mathcal{U}_\sigma(x):
\mathcal{N}(x| m_\sigma^\star,\Sigma_\sigma^\star)
\bigr)
&=
\log\left(
\frac{(2\pi e)^{d/2}d!}
{(d+1)^{(d+1)/2}(d+2)^{d/2}}
\right) \\\forall\sigma&\in\mathcal{S}^{[d]}       
    \end{split}
\end{equation}

Since \(\sum_{\sigma\in\mathcal{S}^{[d]}}w_\sigma=1\), the claim follows.
\end{proof}
%%%%%%%%%%%%%%%%%%%%%%%%%%%%%%%%%%%%%%%%%%%%%%%%%%%%%%%%%%%%%%%%%%%%%%%%%%%%%%%%%%%%%
%%%%%%%%%%%%%%%%%%%%%%%%%%%%%%%%%%%%%%%%%%%%%%%%%%%%%%%%%%%%%%%%%%%%%%%%%%%%%%%%%%%%%
\section{Numerical Examples and Applications}

\begin{example}[$\Pi_{\mathcal{\mathcal{Q}_{\mathcal{N}}}}^{\mathrm{I}}(\mathcal{U}_{\sigma^2}(x))$]
Consider the standard $2$-simplex (an isosceles right triangle):
\begin{equation}
\Delta^2
=
\operatorname{conv}\{0,e_1,e_2\}
\subset\mathbb{R}^2
\end{equation}
with uniform density $\mathcal{U}_{\Delta^2}(x)$. From
\Cref{cor:uniform-pdf-simplex} and
\Cref{prop:dirichlet_mean_var}, its mean and covariance are
\begin{equation}
\begin{split}
m_{\Delta^2}^\star
=
\mathbb{E}_{\mathcal{U}_{\Delta^2}}(X)
=
\frac{1}{3}
\begin{pmatrix}
1\\
1
\end{pmatrix},\quad \Sigma_{\Delta^2}^\star
=
\mathbb{V}_{\mathcal{U}_{\Delta^2}}(X)
=
\frac{1}{36}
\begin{pmatrix}
2 & -1\\
-1 & 2
\end{pmatrix}.
\end{split}
\end{equation}
Hence, by \Cref{prop:min-kl-uniform-standard-simplex-to-gaussian},
the Gaussian I-projection of $\mathcal{U}_{\Delta^2}(x)$ is:
\begin{equation}
\mathcal{N}\left(
x\middle|
m_{\Delta^2}^\star,
\Sigma_{\Delta^2}^\star
\right),
\end{equation}
and the corresponding minimum \ac{KL} divergence is:
\begin{equation}
D_{\mathrm{KL}}\left(
\mathcal{U}_{\Delta^2}(x):
\mathcal{N}(x|m_{\Delta^2}^\star,\Sigma_{\Delta^2}^\star)
\right)
=
\log\left(\frac{\pi e}{3\sqrt{3}}\right).
\end{equation}
Now consider an arbitrary nondegenerate triangle (a simplex with nonvanishing
area)
\begin{equation}
\sigma^2(v_0,v_1,v_2)
=
\operatorname{conv}\{v_0,v_1,v_2\}
\subset\mathbb{R}^2.
\end{equation}
The corresponding affine isomorphism from the standard simplex $\Delta^2$ to
$\sigma^2$ is
\begin{equation}
F:\Delta^2\overset{\mathrm{aff}}{\cong}\sigma^2,
\qquad
x\mapsto Ax+v_0=y,
\end{equation}
where
\begin{equation}
A
=
\begin{bmatrix}
v_1-v_0 & v_2-v_0
\end{bmatrix}.
\end{equation}
Since $\sigma^2$ is nondegenerate, $A$ is invertible. Under this affine transformation, the uniform density $\mathcal{U}_{\Delta^2}$ on $\Delta^2$
is mapped to the uniform density $\mathcal{U}_{\sigma^2}$ on $\sigma^2$. Furthermore, the
Gaussian I-projection transforms according to:
\begin{equation}
m_{\sigma^2}^\star
=
A m_{\Delta^2}^\star+v_0,
\qquad
\Sigma_{\sigma^2}^\star
=
A\Sigma_{\Delta^2}^\star A^\top.
\end{equation}
Consequently,
\begin{equation}
m_{\sigma^2}^\star
=
\frac{v_0+v_1+v_2}{3}.
\end{equation}
Moreover,
\begin{equation}
\Sigma_{\sigma^2}^\star
=
\frac{1}{36}
A
\begin{pmatrix}
2 & -1\\
-1 & 2
\end{pmatrix}
A^\top
=
\frac{1}{12}
\sum_{i=0}^{2}
(v_i-m_\sigma^\star)(v_i-m_\sigma^\star)^\top.
\end{equation}

Finally, by the affine invariance of the \ac{KL} divergence
in \Cref{prop:bijection_kl},
\begin{equation}
\begin{split}
D_{\mathrm{KL}}\left(
\mathcal{U}_{\sigma^2}(y):
\mathcal{N}(y|m_{\sigma^2}^\star,\Sigma_{\sigma^2}^\star)
\right)
=
D_{\mathrm{KL}}\left(
\mathcal{U}_{\Delta^2}(x):
\mathcal{N}(x|m_{\Delta^2}^\star,\Sigma_{\Delta^2}^\star)
\right)
=
\log\left(\frac{\pi e}{3\sqrt{3}}\right).
\end{split}
\end{equation}
Thus, every nondegenerate triangle has the same minimum \ac{KL}
divergence to its Gaussian I-projection, independent of its position,
size, and shape. \Cref{fig:kl-validation} shows random 2-simplices
$\sigma^2$ and Monte Carlo estimates consistent with the analytical minimum
of the \ac{KL} divergence.

\begin{figure*}[htbp]
    \centering
    \includegraphics[width=\textwidth]{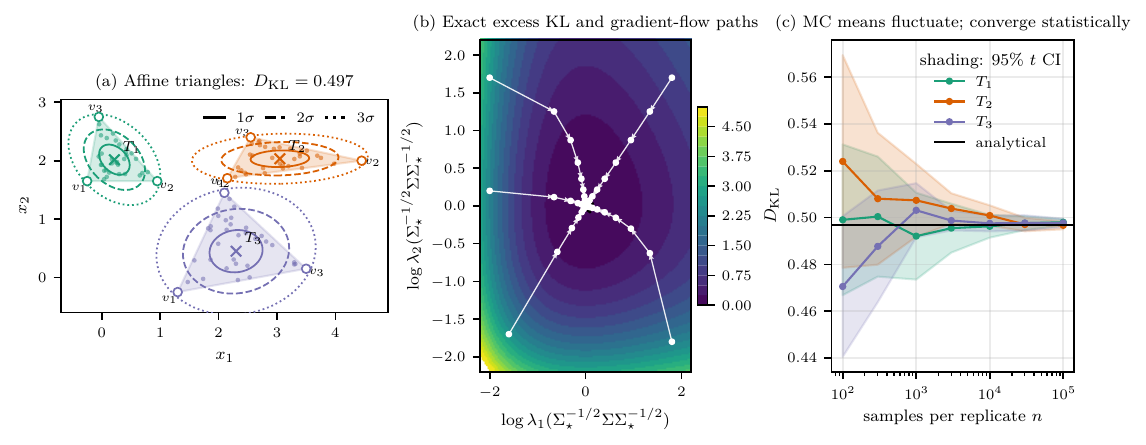}
    \caption{Affine triangles share the proposed analytical simplex based \ac{KL}. Panel b shows the exact excess-KL landscape with gradient-flow paths; panel c shows repeated Monte Carlo estimates and 95\% intervals approaching the analytical value.}
    \label{fig:kl-validation}
\end{figure*}
\end{example}

\begin{example}[Triangle Barycentric Subdivision]
Starting from a single triangle, repeated barycentric subdivision~\cite{elements_of_algebraic_topology_munkres_krantz_parks} produces increasingly fine triangulations of the same domain and, consequently, increasingly detailed \ac{GM} representations. At each of the four finite measured levels in \Cref{fig:barycentric-refinement}, the sampled global forward \ac{KL} divergence and leakage decrease relative to the preceding shown level. These finite observations establish neither general monotonicity under barycentric subdivision nor asymptotic convergence of the \ac{KL} divergence.
\begin{figure}[htbp]
    \centering
    \includegraphics[width=\linewidth]{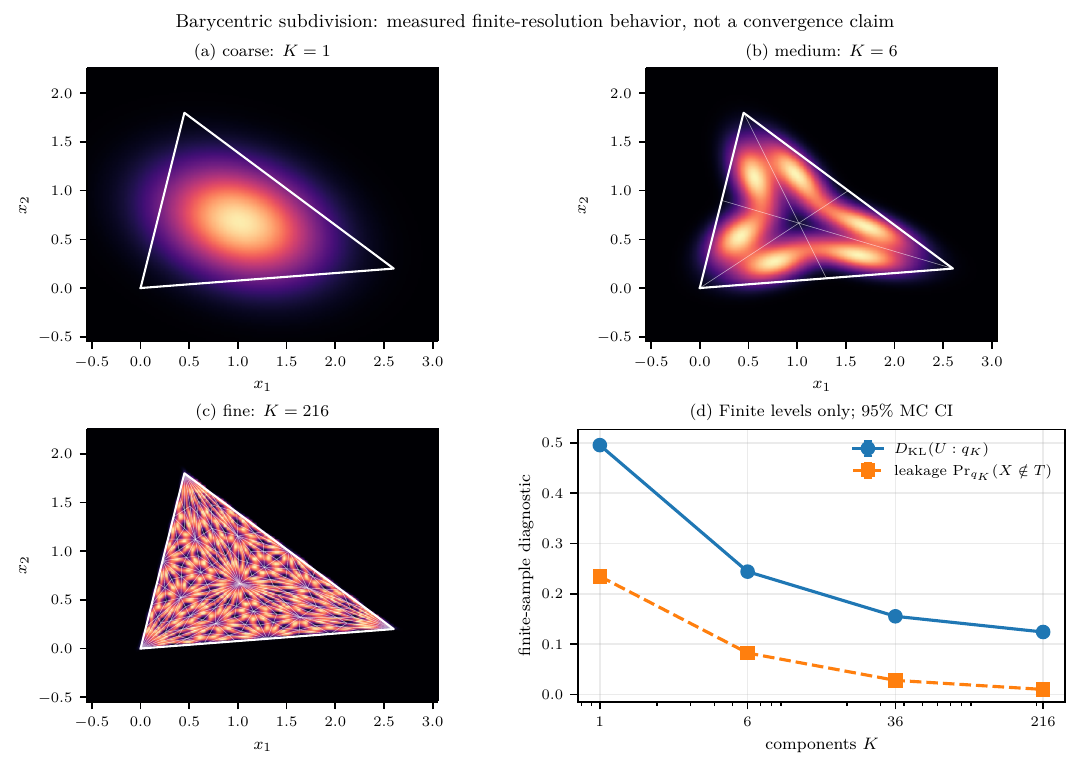}
    \caption{Four finite barycentric-subdivision levels show decreasing sampled forward-\ac{KL} and leakage diagnostics at each displayed level; they do not establish general monotonicity or asymptotic convergence.}
    \label{fig:barycentric-refinement}
\end{figure}
\end{example}

\begin{example}[Polyhedral Refinement]
Starting from a polyhedral domain, increasingly fine triangulations gives \ac{GM} representations with increasing numbers of components. In the displayed run in \Cref{fig:construction-overview}, finer meshes yield lower sampled global forward \ac{KL} divergence for the geometry-derived construction. The held-out forward \ac{KL} of the \ac{EM} fit approaches near zero and then turns upward beginning at $K=256$. Because held-out forward \ac{KL} differs from held-out negative log likelihood only by the target entropy constant, this upturn means that held-out negative log likelihood worsens rather than improves. The behavior is consistent with finite-sample overfitting and/or sensitivity to optimization and regularization, but it does not diagnose a cause. In contrast, the geometry-derived construction has no fitted training sample and hence no train--test overfitting mechanism, although it still incurs approximation and discretization error. At equal component counts, the geometry-derived parameters are obtained directly from the supplied mesh without iterative optimization. The comparison nevertheless has intentionally limited scope: \ac{EM} is one fixed seeded fit using finite training and held-out draws, whereas the geometry timing starts from a supplied mesh and excludes meshing. It is neither a global-optimality claim nor a full benchmark. The finite refinement results also do not establish monotonic decrease or asymptotic convergence in general.
\begin{figure}[htbp]
    \centering
    \includegraphics[width=\linewidth]{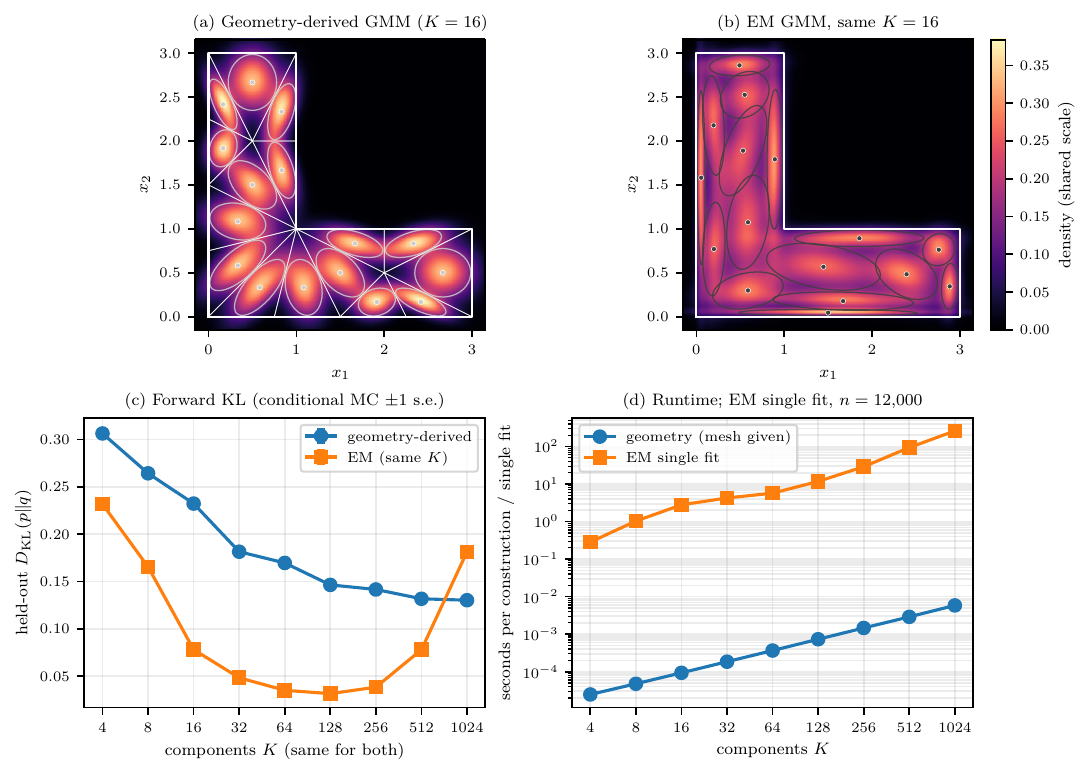}
    \caption{Same-$K$ comparison of the geometry-derived \ac{GM} and one fixed seeded EM fit on finite training and held-out draws. The EM held-out forward \ac{KL} turns upward from $K=256$ in this run. Geometry timing starts from the supplied mesh and excludes meshing; the comparison is not a global-optimality or full-benchmark claim.}
    \label{fig:construction-overview}
\end{figure}
\end{example}
\begin{example}[Map-Aware Fusion]
Let $q_\Omega(x)$ be the geometry-derived \ac{GM} prior, and let
$\ell_r(x)$ and $\ell_c(x)$ denote the robot and camera Gaussian
likelihoods. Under conditional independence of the two measurements given
$x$, the normalized posterior is
\begin{equation}
\begin{split}
q(x\mid z_r,z_c)
&=\frac{q_\Omega(x)\,\ell_r(x)\,\ell_c(x)}{Z},\\
Z&=\int q_\Omega(u)\,\ell_r(u)\,\ell_c(u)\,\mathrm{d}u.
\end{split}
\end{equation}
Multiplying each Gaussian component by the two Gaussian likelihoods in
sequence gives the exact posterior update for the constructed \ac{GM}. The
constrained uniform-domain field shown as a contour in
\Cref{fig:map-aware-fusion} is evaluated on a finite grid and serves only as
a numerical reference. If the sensor correlation is unknown, the conditional
independence factorization is invalid and naive multiplication is not
justified.
    \begin{figure}[htbp]
    \centering
    \includegraphics[width=\linewidth]{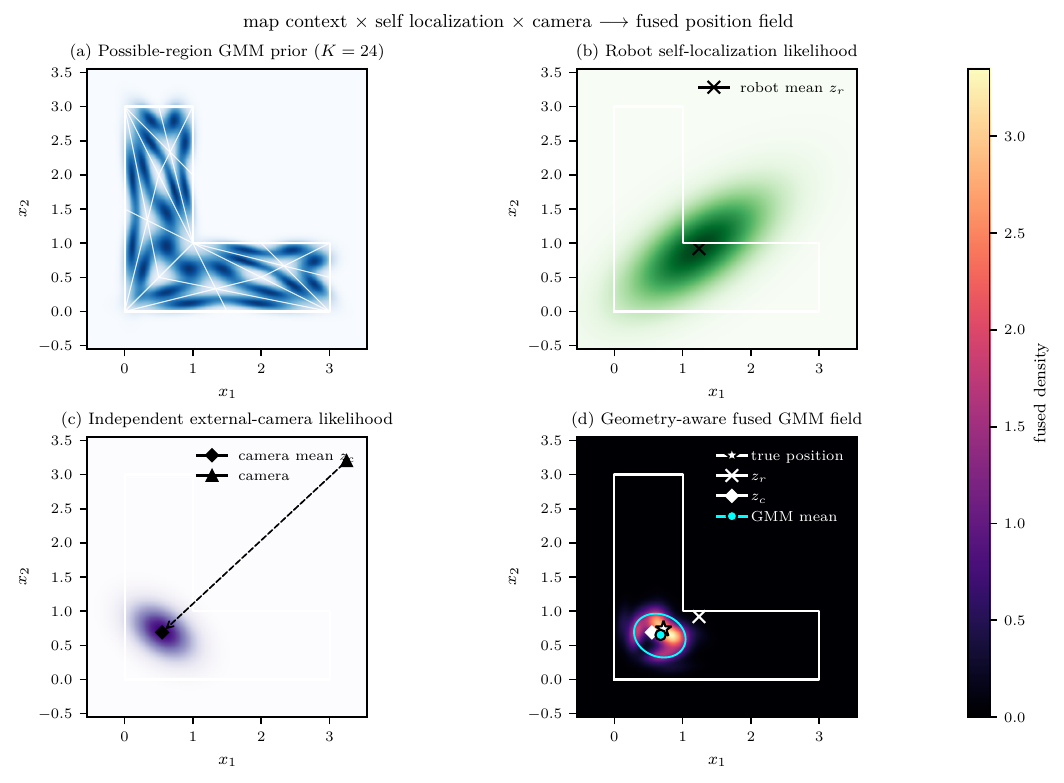}
    \caption{A map prior times conditionally independent robot and camera Gaussian likelihoods gives a sequential closed-form update. The contour is a finite-grid constrained reference; no validity is claimed under unknown sensor correlation.}
    \label{fig:map-aware-fusion}
\end{figure}
\end{example}

\begin{example}[Nut--Bolt Clearance]
Let $e_r$ be the upward radial unit vector. A rigid signed radial translation
by $\delta$, positive upward, transforms the nut density according to:
\begin{equation}
q_{\mathrm{nut},\delta}(x)
=q_{\mathrm{nut}}(x-\delta e_r),
\end{equation}
so every nut-component mean shifts by $+\delta e_r$, while its weight and
covariance remain unchanged. The overlap score is:
\begin{equation}
S(\delta)
=\int_{\mathbb{R}^2}
q_{\mathrm{nut}}(x-\delta e_r)q_{\mathrm{bolt}}(x)\,\mathrm{d}x.
\end{equation}
Writing the two mixtures with parameters
$(w_i^{n},\mu_i^{n},\Sigma_i^{n})$ and
$(w_j^{b},\mu_j^{b},\Sigma_j^{b})$ gives the exact pairwise Gaussian sum:
\begin{equation}
S(\delta)=\sum_{i,j}w_i^{n}w_j^{b}
\mathcal{N}\!\left(
\mu_i^{n}+\delta e_r\middle|\mu_j^{b},
\Sigma_i^{n}+\Sigma_j^{b}
\right).
\end{equation}
These translation and overlap identities are analytic and exact for the
resulting \acp{GM}. The displayed normalized score is relative and dimensionless,
but remains only a soft overlap/interference surrogate: it is not a collision
probability, contact model, ISO analysis, or certification result. A geometry
edit requires remeshing and regeneration of the affected \ac{GM}; a rigid
translation changes only its component means and can be rescored analytically.

\begin{figure*}[htbp]
    \centering
    \includegraphics[width=\textwidth]{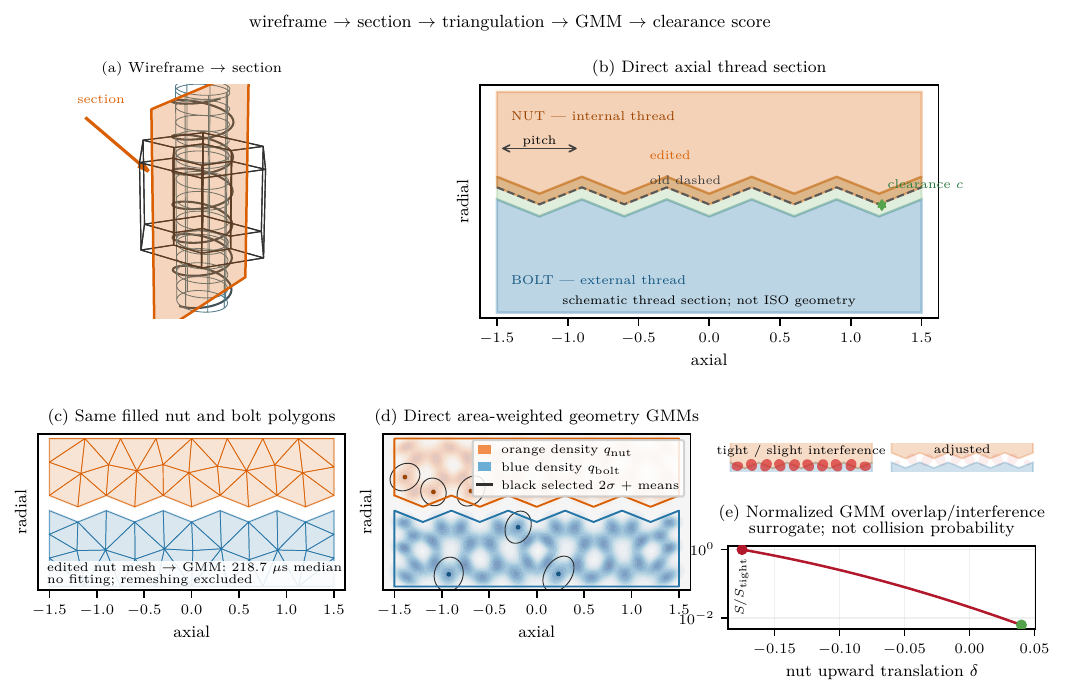}
    \caption{A schematic wireframe is sectioned; the same bounded full-dimensional nut and bolt material polygons are directly triangulated and converted to area-weighted geometry \acp{GM}. Nut-clearance translations update component means and allow analytic overlap rescoring. Normalized overlap/interference is a soft surrogate, not a collision probability; this is not an ISO, contact, or certification analysis.}
    \label{fig:nut-bolt-clearance}
\end{figure*}
\end{example}
%%%%%%%%%%%%%%%%%%%%%%%%%%%%%%%%%%%%%%%%%%%%%%%%%%%%%%%%%%%%%%%%%%%%%%%%%%%%%%%%%%%%%
%%%%%%%%%%%%%%%%%%%%%%%%%%%%%%%%%%%%%%%%%%%%%%%%%%%%%%%%%%%%%%%%%%%%%%%%%%%%%%%%%%%%%
\section{Conclusion}
\label{sec:conclusion}

We considered the problem of representing a uniform distribution on a geometrically described domain $\Omega\subset\mathbb{R}^d$ by a \ac{GM}. Given a finite triangulation of $\Omega$, we proposed a construction that replaces the uniform \ac{PDF} on each $d$-simplex by its \ac{KL}-optimal Gaussian and combines the resulting components according to the volumes of the simplices.

Exploiting the affine isomorphism between simplices and the invariance of the \ac{KL} divergence under bijections, we showed that the corresponding minimum \ac{KL} divergence is independent of the position, size, and shape of the simplex and depends only on the dimension $d$.

By combining the local Gaussian approximations with the volume-weighted representation of the uniform \ac{PDF} on $\Omega$ and using the joint convexity of the \ac{KL} divergence, we obtained a global sharp upper bound. In particular, the resulting \ac{GM} $\mathcal{G}_{\mathcal{S}}(x|g_\mathcal{S})$ satisfies: 
\begin{equation}
D_{\mathrm{KL}}\left(
\mathcal{U}_{\Omega}:\mathcal{G}_{\mathcal{S}}
\right)
\leq
\log\left(
\frac{(2\pi e)^{d/2}d!}
{(d+1)^{(d+1)/2}(d+2)^{d/2}}
\right).
\end{equation}
Thus, the upper bound is independent of the particular geometry and triangulation and depends only on the dimension.

The main advantage of the proposed construction is that it provides an explicit geometry-to-probability mapping without requiring numerical optimization and in closed form. The Gaussian parameters are obtained directly from the vertices of the simplices, while the resulting \ac{GM} preserves the global structure of the original domain. At the same time, the number of the \ac{GM} components is determined by the number of $d$-simplices in the triangulation, so fine triangulations may lead to large \acp{GM}. Moreover, the dimension-dependent bound does not quantify the approximation quality of a particular triangulation. Empirical evidence in~\Cref{fig:barycentric-refinement} suggests that the error decreases, but does not establish asymptotic convergence.

The numerical examples separate validation from application: \Cref{fig:kl-validation,fig:barycentric-refinement,fig:construction-overview} examine the local result, finite refinement behavior, and the deterministic construction relative to a fitted baseline, while \Cref{fig:map-aware-fusion,fig:nut-bolt-clearance} demonstrate closed-form Gaussian fusion, rigid translation, and analytic overlap evaluation using the resulting \acp{GM}.

This work provides an explicit construction and a dimension-dependent upper bound in the \ac{KL} sense for \ac{GM} representations of triangulated domains. Future work could investigate adaptive triangulations and component reduction methods to obtain more compact representations while controlling the approximation error. Extensions to more general geometric objects, including lower-dimensional manifolds embedded in higher-dimensional spaces, are also of interest.

On the application side, we envision ``Probabilistic Physics Using \ac{GM}'' as a natural next step in the \ac{GM} based geometric formulation of physical systems. Geometry has provided a remarkably successful language for formulating physical laws and capturing the underlying structure of physical systems, from geometric formulations of classical mechanics to modern theories of dynamics and field physics~\cite{marsden_geometric_mechanics,marsden_symmetry_mechanics,arnold_mathematical_methods_of_classical_mechanics,frankel_geometry_of_physics}. Probability, in turn, provides a principled language for representing uncertainty, variability, and incomplete knowledge, forming the foundation of modern state estimation and sensor fusion. Building on this probabilistic representation, stochastic and optimal control provide a natural framework for reasoning and decision-making under uncertainty~\cite{fleming_rishel_deterministic_and_stochastic_optimal_control}, opening the possibility of methods that jointly learn, model, estimate, and control complex physical systems.
%%%%%%%%%%%%%%%%%%%%%%%%%%%%%%%%%%%%%%%%%%%%%%%%%%%%%%%%%%%%%%%%%%%%%%%%%%%%%%%%%%%%%
%%%%%%%%%%%%%%%%%%%%%%%%%%%%%%%%%%%%%%%%%%%%%%%%%%%%%%%%%%%%%%%%%%%%%%%%%%%%%%%%%%%%%
\section{Acknowledgment}
This work was funded by Project SFB~1574 ``A Circular Factory for the Perpetual Product'', Deutsche Forschungsgemeinschaft (DFG), Project ID 471687386.
%%%%%%%%%%%%%%%%%%%%%%%%%%%%%%%%%%%%%%%%%%%%%%%%%%%%%%%%%%%%%%%%%%%%%%%%%%%%%%%%%%%%%
%%%%%%%%%%%%%%%%%%%%%%%%%%%%%%%%%%%%%%%%%%%%%%%%%%%%%%%%%%%%%%%%%%%%%%%%%%%%%%%%%%%%%
\printbibliography
\end{document}